\documentclass[aps,prb,reprint,superscriptaddress,showpacs]{revtex4-1}
\usepackage{bbm,amsmath,amsfonts,amssymb,indentfirst,syntonly,graphicx}
\usepackage{amsmath,amssymb,amsthm}
\usepackage{latexsym,graphicx,bbm}
\newcommand{\beq}{\begin{equation}}
\newcommand{\eeq}{\end{equation}}
\newcommand{\p}{\partial}
\newcommand{\Tr}{{\rm Tr}}

\begin{document}

\title{Non-Abelian vortices in the emergent $U(2)$ gauge theory of the Hubbard model}
\date{\today}
\author{Qiu-Hong Huo}
\affiliation{Laboratory of Thin Film Materials, College of Materials Science and  Engineering,
 Beijing University of Technology, 100124 Beijing , China}
 \author{Yunguo Jiang}
 \email[]{jiangyg@ihep.ac.cn}
 \affiliation{Institute of High Energy Physics, Chinese Academy of Sciences, 100049 Beijing, China}
 \affiliation{Theoretical Physics Center for Science Facilities, Chinese Academy of Sciences, 100049 Beijing,China}
\author{Ru-Zhi Wang}
 \email[]{wrz@bjut.edu.cn}
\affiliation{Laboratory of Thin Film Materials, College of Materials Science and  Engineering,
 Beijing University of Technology, 100124 Beijing , China}
 \author{Hui Yan}
  \email[]{hyan@bjut.edu.cn}
 \affiliation{Laboratory of Thin Film Materials, College of Materials Science and  Engineering,
 Beijing University of Technology, 100124 Beijing , China}

\begin{abstract}
By the spin-fermion formula, the Hubbard model on the honeycomb lattice is represented by a $U(2)$ gauge
 theory in the mean field method, non-Abelian vortex solutions are constructed based on this theory. The quantization
 condition shows that the magnetic flux  quanta are half-integer. There are $2k$ bosonic zero modes  for $k$ winding vortices. For the fermions, there are 2 zero energy states (ZESs) corresponding to the single elementary vortex. In the vortex core and on the edge, the system  are in  the semi-metal phase with a spin gap and
 in the insulator phase with N\'eel order phase, and can be mapped to the superconductor in class A and CI, respectively.
\end{abstract}
\maketitle

In 2+1 dimensional systems, electrons moving on the lattice can be well described by the Hubbard model.
It is a long-standing research topic due to the  richness of its phase diagram and its possible explanation of
high-$T_c$ superconducting. The Hubbard model on the honeycomb lattice is also suitable to study  novel materials like the graphene,
which has a great potential for applications. Previously, Hou, Chamon and Mudry (HCM) discussed the Abelian vortex
in graphene-like structures \cite{Hou:2006qc}, where the vortex configuration of the order parameter is caused by the distortion on the  Kekul\'e texture. Herbut used the Hubbard model to consider vortices in the background N\'eel order, and found two orthogonal ZESs of
   fermions at the vortex core \cite{Herbut:2007zz}.
Nevertheless, the vortex configurations in the above cases are of Abelian type.

In recent years, vortices of non-Abelian type have been discovered in ${\cal N}=2$ supersymmetric QCD (SQCD)
\cite{Hanany:2003hp,Shifman:2004dr,Auzzi:2003fs}, which is indeed a breakthrough since
the Abelian Abrikosov-Nielsen-Olesen (ANO) vortex has been constructed \cite{Nielsen:1973hp}.
The similarity of the emergent gauge theory (EGT) of the Hubbard model and the bosonic truncation of
 SQCD motivates us to construct the non-Abelian vortices in this theory.
In this work, we present a complete investigation of the EGT of the Hubbard model in the
 low energy limit \cite{Sachdev:2010uz}. By the mean-field theory method, we construct the non-Abelian vortex solutions in the theory. Non-Abelian vortices have non-trivial orientational
zero-modes, which induce net interactions between multi-vortices \cite{Eto:2011pj}.
  We hope that the non-Abelian vortices enable us to understand novel phenomenons
in condensed matter physics, especially the unconventional superconductor.

 The honeycomb lattice  has a special property that the occupied and empty
 states meet at two Fermi points in  momentum space.  Due to the  bipartite property of the Honeycomb
  lattice, one can introduce three Pauli matrices $\sigma^a$ , $\tau^a$, and $\rho^a$ to act on  the
   spin, the sublattice (pseudospin) and the valley spaces, respectively. The standard Hamiltonian  of the Hubbard
   model is written as $H_t+H_U$, where $H_t$ describes the  hopping of electrons, and $H_{U}$
   describes the repulsive interaction of electrons. Expanding $H_t$  near the Fermi points,
   one obtains \cite{Sachdev:2009cp}
\beq H_t=\frac{3t}{8\pi^2} \int d^2k C^{\dag}_{p i
  s}(\mathbf{k}) \big(\tau^y_{pq}k_x+
\tau^x_{pq} \rho^{z} _{ij} k_y  \big) C_{q j s}(\mathbf{k}), \label{eq:Hamiltonian} \eeq
where $t$ is the hopping constant. The index $s=(\uparrow, \downarrow)$ denotes the spin,
$p,q=(A,B)$ denote the sublattice, and $i,j=(1,2)$ denote the valley. With $H_U$ term, the Hubbard model has
a very rich phase diagram.

The emergent gauge theory kicks in an elegant way to unify phases of the Hubbard model.
% It is tempting to investigate the emergent gauge symmetry of the Hubbard model. %Performing the Hubbard-Stratonovic transformation, the Hubbard model was converted to the Gross-Neveu-Yukawa (GNY) theory in the continuum limit.
 %Based on that, an emergent $U(1)$ gauge theory was constructed \cite{Sachdev:2009cp}.
  Hermele constructed an $SU(2)$ gauge theory for the Hubbard model on the honeycomb lattice, where the  electron operators were written in the slave-rotor formulation, and an SU(2) algebra spin liquid (ASL) is found on the insulator side of Mott transition \cite{Hermele:2007pp}.
 Sachdev et al. constructed an $SU(2)$ gauge theory of the Hubbard model in a different way \cite{Sachdev:2010uz,Sachdev:2009cp}, which will be presented in the following. By making use of the spin-fermion formula, the electron operators can be decomposed as
\beq \begin{pmatrix} c_{\uparrow} \\
                     c_{\downarrow} \end{pmatrix}= \begin{pmatrix}
   R_{11} & R_{12} \\
                    R_{21} &   R_{22} \end{pmatrix}
 \begin{pmatrix} \psi_{+} \\
                     \psi_{-} \end{pmatrix} \equiv R \, \Psi\ , \label{eq:decom}
\eeq
where $R$ denotes the spin density wave (SDW) order.  $\psi_{\pm}$ are the spinless
fermion operators,  the indices $\pm$ measure the spin-projection along the local spin reference axis \cite{Sachdev:2009cp}.
The $R$ fields can also be interpreted as the "auxiliary boson" in the formulation of Kotliar and Ruckenstein (KR),
which represents four boson operators corresponding to four atomic states per site \cite{kr1986}.
  In Ref \onlinecite{Sachdev:2010uz}, a Higgs-like potential has been
  constructed for $R$, which leads to symmetry breaking and is necessary for vortex solutions. In this manner,
  $R$ is generalized to an invertible complex $2 \times 2$  matrix,
  which represents a generic spin operator for arbitrary filling \cite{035106}.

Let us show how the hidden gauge symmetry emerges from the spin-fermion formula.
First, the $R$ matrix can be rotated by an $SU(2)$ global spin rotation from left
\beq R \to V\,R, \qquad \Psi \to \Psi \, \qquad C \, \to V\, C\ . \eeq
As this action is global, the spin rotation will be regarded as the "flavor" symmetry.
Secondly, there is a local gauge redundancy, the theory is invariant under the gauge transformations
\beq R \to R\,U^{\dag} \ , \qquad  \Psi \to U \,\Psi \ , \qquad C \to C \ , \eeq
where $U$ is an element of a gauge group. Sachdev et al. consider  $U \in SU(2)$ , here we argue that the system is
 invariant  under $U(2)$, and the theory is similar to the Weinberg-Salam (WS) electro-weak
 theory, but limited to $2+1$ dimensions.
This extra $U(1)$ gauge group is the key ingredient for the non-Abelian vortices, because the homotopy group
is non-trivial for  $U(1)\times SU(2)$, i.e.,
\begin{equation} \label{eq:topological}
\pi_1 \left(\frac{U(1)\times SU(2)}{\mathbbm{Z}_2}\right) = \mathbbm{Z}.    \
\end{equation}
Instead, when the $SU(2)$ gauge group is considered, only $\mathbbm{Z}_2$ vortex can be constructed,
%since $ \pi_1 \left( SU(2) \right) = \mathbbm{Z}_2$,
  which is genuinely of Abelian type \cite{hep-th/0106175}.
 %The $U(1)$ gauge group may have the different coupling constant with the $SU(2)$ part, but this will not
%impair our analysis, and only  introduce diverse  Bogomol¡¯nyi-Prasad-Sommerfield (BPS) equations.
% Here we expose that they are equal for simplicity.
%The $U(1)$ gauge field can describe the interaction between the SDW and spinless quasiparticles,
%the SDW-photon-quasiparticle interaction may play an important role in high-$T_c$
% superconductor.

The Lagrangian of the  emergent $U(2)$ gauge theory of the Hubbard model
 reads \cite{Sachdev:2010uz}
\begin{align} {\cal L}= &\bar{\Psi} \gamma_{\mu}(\partial_{\mu}+i \sigma^{\alpha} A_{\mu}^{\alpha})
\Psi-\lambda \Phi^a\bar{\Psi}\rho^z\sigma^a \Psi \nonumber \\
    &+ \frac{1}{2}\big[ (\p_{\mu}\Phi^a-2\epsilon_{abc} A_{\mu}^b\Phi^c)^2 \big]+s
    (\Phi^a)^2+ u(\Phi^a)^4 \nonumber \\
 &+\Tr \big[ {\cal D}_{\mu}R({\cal D}_{\mu}R)^{\dag }\big]+\tilde{s} \Tr
 (R^{\dag}R)+\tilde{u}[\Tr(R^{\dag}R)]^2 \ , \label{eq:lagrangian}\end{align}
where  the covariant derivative is defined to be $ {\cal D}_{\mu}R \equiv \p_{\mu}R -iA^{\alpha}_{\mu}R \sigma^{\alpha}$.
The index $\alpha$ runs from $0$ to $3$, meanwhile  $a,b$ take value as $a,b=1,\cdots,3$ in
 the following.  "$0$" stands for the Abelian $U(1)$ part of the gauge group, and $a,b$ stand for the non-Abelian
  $SU(2)$ part. The representation of the algebra is that $\sigma^0=\mathbf{1}_2$, $\sigma^a$ is the Pauli matrix,
  which follows that ${\rm Tr} [\sigma^{\alpha}\sigma^{\beta}]=2 \delta^{\alpha \beta}$. Hence $\Phi^a$ transforms
  as the triplet of gauge  $SU(2)$, $\rho^z$ is the third component of the Pauli matrices which acts on the  valley space, see Eq.(\ref{eq:Hamiltonian}).
  The theory has a set of symmetries ${\rm U(1)_{g}\otimes SU(2)_{g}\otimes SU(2)_{s}}$, the transformation properties
   of the matter contents $\Psi$, $\Phi^a$ and $R$ are listed in Table \ref{ta:transform}.

\begin{table}
\begin{center}
\begin{tabular}{|c||c|c|c|}
     \hline Matter content   & ${\rm U(1)_{gauge}} $& ${\rm SU(2)_{gauge}}$ & ${\rm  SU(2)_{spin}} $ \\ \hline
        $\Psi $& 1 & 2 & 1\\ \hline
                $\Phi^a$ & 0 & 3 & 1 \\ \hline
                  $  R $& 1 & $\bar{2}$ & 2  \\ \hline
                    \end{tabular}
                    \caption{The transformation properties of the fields.}
                    \label{ta:transform}
\end{center}
\end{table}

\begingroup
\squeezetable
 \begin{table}
\begin{center}
\begin{tabular}{|c||c|c|}
     \hline   & $\langle \Phi^a \rangle $ =0 &   $\langle \Phi^a \rangle  \ne 0 $ \\ \hline
        $\langle R \rangle = 0 $ & {\rm Semi-metal  } & {\rm
        Insulator with VBS order} \\ \hline
        $\langle R \rangle \ne 0$ & {\rm Semi-metal with a spin gap} & {\rm Insulator with Neel order} \\ \hline
                                   \end{tabular}
                    \caption{The phase diagram of the system with $U(2)$ gauge theory \cite{Sachdev:2009cp}, in which
                    VBS is the abbreviation of valence bond solid.}
                    \label{ta:Phase2}
\end{center}
\end{table}
\endgroup

The  $s$, $u$  and tildes of them are the coupling constants which tune the couplings of $R, \Phi^a$.
Four different phases are unified in the Lagrangian of Eq.(\ref{eq:lagrangian}). They are classified by whether the
vacuum expectation values (VEV) of the $R$ and $\Phi^a$ are zero or not \cite{Sachdev:2010uz}, see  Table \ref{ta:Phase2}. When the adjoint scalar $\Phi^a$ is condensed at a energy scale  $v_1^2 \equiv -s/2u$,
 the gauge symmetry breaking patten is $SU(2) \to U(1)$. the Lagrangian in Eq.(\ref{eq:lagrangian})
 reduces to the $U(1)$ gauge theory, where monopole events (or hedgehog) occur \cite{Senthil:2004cp}.
  When $\Phi^a$ acquires a VEV, the Yukawa coupling term $-\lambda \Phi^a\bar{\Psi}\rho^z\sigma^a \Psi$ will become a mass term for $\Psi$,  the effective mass  is  $m_{\Psi}=\sqrt{2\lambda v_1}$.
We decouple $\Phi^a$ by  a hierarchical symmetry breaking, i.e. $v_1 \gg v_2$, where $v_2$ is the VEV of $R$.
   The spectrum of $\Phi^a$ will not appear below the low energy scale $v_2$.

$R$ potential in Eq.(\ref{eq:lagrangian}) is generic, we choose a special potential of $R$ in the following, which is written as
 \begin{equation} \label{eq:rpotential}
 {\cal V}(R)=\tilde{u} {\rm Tr} \left[ \big(R^{\dag}R - \frac{1}{2}v^2_2 \mathbf{1}_2 \big)^2\right],
 \end{equation}
where $v_2^2 = -\tilde{s}/\tilde{u}$.  Up to gauge rotations,
 the VEV of $R$ is  $\langle R \rangle =v_2 /\sqrt{2} \mathbf{1}_2$,
which is called the "color-flavor" locked (CFL) phase. In the Higgs phase, the masses of the two flavors (spin indices) are the same.
 In the CFL phase, the $U(1)\times SU(2)$ gauge symmetry is broken. Nevertheless, there is a diagonal invariant rotation, that is
\beq
 V_{f} \langle R \rangle U_{c}^{\dag}= \langle R \rangle , \qquad \,V_f, \,
 U_c^{\dag}  \in SU(2) . \label{eq:colorflavor}
\eeq
 This means the spin rotation of $R$ can be equally done by the gauge rotation in the vacuum.
  So a combined color-flavor $SU(2)_{c+f}$ global symmetry remains.

With all the criteria above, at an energy scale below $v_1$, the effect action  can be
written as
 \begin{align} {\cal S}= &{\cal S}_{\Psi}+{\cal S_{R}}, \\
 {\cal S}_{\Psi}=&\int d^3x \{ \bar{\Psi} \gamma_{\mu}{\cal D}_{\mu}
\Psi+\frac{1}{2}m_{\Psi}^2\bar{\Psi}\rho^z\sigma^a \Psi \},  \\
    {\cal S_{R}}= &\int d^3x \, \Tr  \{ \frac{1}{4g^2}|F_{\mu \nu}|^2+
    |{\cal D}_{\mu}R|^2+{\cal V}(R) \} ,     \label{eq:lagrangian2}
\end{align}
where $F_{\mu \nu}\equiv i[D_{\mu}, D_{\nu}]$, and $g$ is the coupling constant of gauge fields. The Yang-Mills
term is added by the consideration of relativistic dynamics.
The static vortex configuration of the system is the minimal energy bound.

The action in Eq.(\ref{eq:lagrangian2}) is the non-Abelian generalization of
 the Abelian Higgs model. Setting $\beta \equiv 2\tilde{u}/g^2$, which is the
 Landau-Ginzburg parameter classifying the types of superconductor.
   We will work in the BPS limit  $\beta=1$, which means $|\tilde{u}|=g^2/2$.
 Performing the  Bogomol'nyi completion, one obtains the BPS equations of the system
\begin{align} \label{eq:BPS}
\bar{\cal D}R=&0, \\
F_{12}-g^2\big( R^{\dag}R - \frac{v^2_2}{2} \mathbf{1}_2\big)=&0, \label{eq:BPS2}
\end{align}
where $2\bar{\cal D}={\cal D}_1+i {\cal D}_2$ and $z=x_1+i x_2$ is the standard
 complex coordinate. The energy bound of the system is
\beq {\cal E} \ge -\int d^2x \, \Tr( F_{12} v_2^2) = -2\pi v_2^2 k, \qquad k \in \mathbbm{Z}. \eeq
 The generators of $SU(2)$ are traceless, so it does not contribute to the
  bound energy. Meanwhile, the $U(1)$  part does contribute, and the minimal
  energy bound is the  quantized magnetic flux. Notice that our winding number
  is $k/2$, which is a half-integer  \cite{arXiv:0802.1020}. The  normalization condition,
  $\Tr (\sigma^0)=2$, cancels the "2" factor  on the denominator of the winding number.
  % The direction of the flux is orthogonal  to the complex plane.
     This can explain naturally the half-integer magnetic flux quanta
  found in certain superconductor materials.

Choosing the minimal winding,
the ANO-like Ansatz can be embedded in the upper left-hand corner in $R$ , i.e,
\beq R=\begin{pmatrix}
      e^{i \theta} f_1(r) & 0\\
      0 & f_2(r)
       \end{pmatrix}, \label{eq:Ransatz}
       \eeq
 which is named the ($1,0$) vortex. Similarly, the winding term can also be embedded in
  the lower right-hand corner, which is named  the  ($0,1$) vortex. Both cases have
   degenerate energy. Here  $f_1(r)$ and $f_2(r)$ are the profile functions, the boundary
   conditions for them are $ f_{1}(\infty)=f_2(\infty)=v_2/ \sqrt{2}$,  $f_1(0)=0$ and  $ \partial_r f_2(0)=0$.
Requiring ${\cal D}_i R \to 0$ when $r \to \infty$, the Ansatz for the gauge field $A_i$ is written as
\beq A_i=-\frac{1}{2} \epsilon_{ij}\frac{x_j}{r^2} \left[\left(1-g_1(r)\right)\mathbf{1}_2+
 \left(1-g_2(r)\right)\sigma^3 \right]. \eeq
The profiles $g_{1,2}$  satisfy the boundary conditions $ g_{1,2}(\infty)=0$ and $g_{1,2}(0)=1$.
One can see that it is the combination of $U(1)$ and $SU(2)$ which cancels the divergence
 of $\partial_i R$. Substituting the Ansatz into the Eq.(\ref{eq:BPS}) and (\ref{eq:BPS2}), one has the BPS equations for the profiles
\begin{align}
\frac{d}{dr}f_1&-\frac{f_1}{r}(g_1+g_2)=0,\;\frac{1}{r}\frac{d}{dr}g_1-g^2(f_1^2+f_2^2-v_2^2)=0,  \nonumber \\
  \frac{d}{dr}f_2&-\frac{f_2}{r}(g_1-g_2)=0,\;\frac{1}{r}\frac{d}{dr}g_2-g^2(f_1^2-f_2^2)=0.
\end{align}
These equations can be solved  numerically.

In Eq.(\ref{eq:Ransatz}), the configuration of the ($1,0$) vortex breaks the global $SU(2)_{c+f}$ color-flavor symmetry down to $U(1)_{c+f} \in SU(2)_{c+f}$ , the Nambu-Goldstone-like modes arise, which are named
 as the internal orientational zero modes, expressed as \cite{Shifman:2004dr}
\beq \mathbbm{C}P^1 \simeq \frac{SU(2)_{c+f}}{U(1)_{c+f}}. \eeq
Other solutions can be generated by "color-flavor" transformations, i.e., $R \to U_{c+f}
 R U_{c+f}^{\dag}$, $A_i \to U_{c+f} A_i U_{c+f}^{\dag}$, where $U_{c+f}$ parameterizes the
 $\mathbbm{C}P^1$. All solutions have the same energy and the same boundary condition up to a regular gauge transformation.  Thus,  we have a moduli space. One can use the index theorem to calculate the dimension
   of the moduli space, which is found to be ${\cal I}= N_c \, N_f\,k/2=2k$ \cite{Hanany:2003hp}.
   These bosonic zero modes stand for the remanent excitation of order parameter, and are massive modes in the bulk
    regime (where $R$ is in vacuum). Therefore, they are localized in the vortex core \cite{arXiv:1007.2116}.
In the low energy, the dynamics of the zero modes can be well described by the $\mathbbm{C}P^1$ sigma model \cite{Shifman:2004dr,Auzzi:2003fs}.
Our analysis above is semi-classical. When quantum mechanism is considered, the $\mathbbm{C}P^1$ sigma model
will develop two vacua corresponding to the (1,0) and (0,1) elementary vortices, respectively \cite{Shifman:2004dr}.

Next, we will consider how many ZESs  for the $8$  components Dirac fermions.
For Abelian vortices, $k$ ZESs of fermions in the vortex-fermion system  correspond to the $k$-vortex background \cite{Jackiw:1981ee}. The HCM model discussed the Abelian vortex in  Kekul\'e order parameters, see Eq.(3) in Ref.\onlinecite{Hou:2006qc}. When the spin indices are summed,  two ZESs are expected for 8 components Dirac fermions. Turning on the $H_U$ term, Herbut also found two orthogonal ZESs at the center of the vortex  for the Hubbard model on the honeycomb  lattice
  \cite{Herbut:2007zz}, but the vortex configuration is in the  N\'eel order $N^a$, which exists in the spin space.
The coincidence of the two instances is due to the fact that the N\'eel order $\vec{N}=(N_1,N_2,0)$ in $H_U$
 will turn into an Kekul\'e-like order $\Delta$, i.e., $\Delta=N_1+iN_2$, when the
 the Hamiltonian is expressed explicitly by using the conventional 4 component Dirac spinor $(u_b, u_a, v_a,v_b)
^{T}$ as in Ref.\onlinecite{Hou:2006qc}. For our non-Abelian vortices, one of the two elementary vortex configurations
(1,0) and (0,1) survives after considering the quantum effects, and can be set as the background vortex.
Essentially, the (1,0) or (0,1) vortex is of Abelian type.
Since $R$ can represent the N\'eel order by the relation $N^a \tau^a=R \tau^3 R^{\dag}$,
 analogous zero energy Dirac equations in Herbut's case can be given
for Hamiltonian in Eq.(\ref{eq:Hamiltonian}).
So we expect two  ZESs for the (1,0) or (0,1) vortex in the extended Hubbard model, and $2k$ ZESs for the $k$ winding vortices.

It is still an open question how the bosonic orientational zero modes interact with the fermions in the vortex core,
a self-consistent way to treat this question is to draw on the Bogoliubov-de Gennes (B-dG) equation  \cite{Yasui:2010yw}.
In the B-dG equations, the fermion zero modes depend non-trivially on the winding term, it can be inferred that the degree
freedom of the bosonic zero modes will not affect that of the fermion zero modes. It is shown that  the fermion zero
modes is well-localized in the vortex core \cite{Jackiw:1981ee,Yasui:2010yw}. In Ref.\onlinecite{Yasui:2010yw}, singlet and
triplet states of the unbroken symmetry are found for the non-Abelian vortices in high density QCD, where the $SU(3)_{c+L+R}$
symmetry is broken to $SU(2)_{c+L+R} \times U(1)_{c+L+R}$. In our model, only $U(1)_{c+f}$ symmetry group exists after symmetry
breaking, one singlet state can be constructed. We leave the B-dG equations of the extended Hubbard model for future work.

It is interesting to notice that the vortex core and the vortex edge are in quite different phases.
Far away from the vortex core,  the gauge bosons $A_{\mu}^{\alpha}$ obtain the mass $\sim 2 g v_2$.
The magnetic flux $F_{12} \propto 1/r^2$ asymptotes to zero.
The non-existence of magnetic fields preserves the time reversal symmetry (TRS).
With the set-up $\langle \Phi^a \rangle \ne 0$, $\Phi^a$ will turn into a N\'eel order parameter. The system is
in the insulator phase  with the antiferromagnetism (N\'eel) order \cite{Sachdev:2010uz}.
 If $\langle \Phi^a \rangle= 0$, the system is in the semi-metal phase.
In the CFL vacuum, the spin rotation symmetry (SRS) is respected.
The insulator with both TRS and SRS  is associated with superconductor in class
 CI according to Altland-Zirnbauer (AZ) classification
 \cite{Roy:2010pp,PhysRevB.55.1142}.

In the vortex core, the physics is rich. $\langle \Phi^a \rangle \ne 0$ means
that the $SU(2)$ gauge symmetry is broken to $U(1) \in SU(2)$ \cite{Sachdev:2010uz}.
The half quanta of magnetic flux pass through the core, so the TRS is broken.
Considering the (1,0) vortex as the background, the boundary conditions of the profiles
 lead to $R= {\rm diag}(0, f_2(0))$ when $r \to 0$, where $f_2(0)=\eta$ is a normalized constant.
 This indicates that the $U(2)$ gauge and $SU(2)$ spin symmetry are not
 completely restored in the vortex core.
 Evidently, the $SU(2)$ SRS is not only broken along the $\sigma^3$ axis.
 This can be mapped to the superconductor in class A \cite{prb195}. A spin discrepancy occurs,
 i.e., $C_{\uparrow}=0, C_{\downarrow}=\eta \psi_{-}$. The spin up fermions are suppressed, one can expect
 that the magnetism occurs in the core.  If we take the (0,1) vortex, the magnetism will inverse the direction.
 In the presence of one pair of them, we will have a magnetic dipole in the system.
   Recall that  our gauge symmetry is $U(2)$ , an $U(1) \times
  U(1)$ gauge theory is not broken. With two spin flavors, the system becomes an $U(1) \times
  U(1)$ two-gap superconductor \cite{cond-mat/0201547}.
  $\Psi$ turns to a massive fermion when $\langle \Phi^a \rangle \ne 0$. Integrating them out,
   ${\cal S}_{\Psi}$ will induce an effective non-Abelian Chern-Simons (CS) action \cite{arXiv:1010.3724},
   i.e., ${\cal S}_{CS}= \frac{1}{4\pi} \int d^3x \epsilon^{\mu \nu \lambda}[A_{\mu}^a
   \partial_{\nu} A_{\lambda}^a+ \frac{1}{3} \epsilon^{abc}A_{\mu}^a A_{\nu}^b A_{\lambda}^c]$.
    Combining with ${\cal S}_{R}$, the theory will turn into non-Abelian Chern-Simons-Higgs theory.
    The winding number of the case is found to be $k/n_0$ and $(k+ q)/ n_0$ ($k$ is an integer,
     $n_0$ is the center of the gauge group, $q$ is a real number) for topological and non-topological vortices,
   respectively \cite{Gudnason:2009ut}. If $\langle \Phi^a \rangle =0$, the $\Psi$ fermions are massless. $R$ has a spin gap, so the system are in the semi-metal with a spin gap phase.

The construction of non-Abelian vortices depends on the spin-fermion formula. In our analysis, we do not consider
the physics in the valley and sublattice spaces.  Thus, this decomposition can be  applied to other types of lattices. Our discussion also holds for the 3 dimensional physics with $x_3$ as the direction of the vortex string.

In conclusion, we show that there is an local gauge redundancy in the emergent gauge theory of the Hubbard model,
and this gauge theory is $U(2)$. Based on the  effective Lagrangian in Eq.(\ref{eq:lagrangian}),
we construct the non-Abelian vortex solutions , and show that there are $2k$ orientational zero modes for $k$-winding vortex. For the elementary vortices, there are $2k$  ZESs for the fermions. Far away from the vortex core, the system is in the insulator phase with N\'eel order. In the vortex core, the system are in the semi-metal phase with a spin gap.

\begin{acknowledgments}
We would like to than Jarah Evslin for valuable discussions. The work of Y. Jiang was funded by the National Natural Science Fund of China under Grant No. 10875129 and No. 11075166.
\end{acknowledgments}


\begin{thebibliography}{99}




\bibitem{Hou:2006qc}
  C.~Y.~Hou, C.~Chamon, and C.~Mudry,
  %``Electron fractionalization in two-dimensional graphene-like structures,''
  Phys.\ Rev.\ Lett.\  {\bf 98}, 186809 (2007).
  %[cond-mat/0609740 [cond-mat.mes-hall]].

  %\cite{Herbut:2007zz}
\bibitem{Herbut:2007zz}
  I.~F.~Herbut,
  %``Zero-energy states and fragmentation of spin in the easy-plane antiferromagnet on honeycomb lattice,''
  Phys.\ Rev.\ Lett.\  {\bf 99}, 206404 (2007) .
 % [arXiv:0704.2234 [cond-mat.str-el]].

%\cite{Hanany:2003hp}
\bibitem{Hanany:2003hp}
  A.~Hanany, and D.~Tong,
  %``Vortices, instantons and branes,''
  JHEP {\bf 0307}, 037 (2003).
 % [hep-th/0306150].
 %\cite{Shifman:2004dr}
\bibitem{Shifman:2004dr}
  M.~Shifman, and A.~Yung,
  %``NonAbelian string junctions as confined monopoles,''
  Phys.\ Rev.\ D {\bf  70}, 045004 (2004).
 % [hep-th/0403149].Auzzi:2003fs

\bibitem{Auzzi:2003fs}
  R.~Auzzi, S.~Bolognesi, J.~Evslin, K.~Konishi, and  A.~Yung,
  %``NonAbelian superconductors: Vortices and confinement in N=2 SQCD,''
  Nucl.\ Phys.\  B {\bf  673}, 187 (2003).
%  [hep-th/0307287].

\bibitem{Nielsen:1973hp}
 H. B.~Nielsen and P.~Olesen,% ¡°Vortex-line models for dual strings,¡±
 Nucl.\ Phys.\ B {\bf 61}, 45 (1973) .
\bibitem{Sachdev:2010uz}
  S.~Sachdev,
  %``The landscape of the Hubbard model,''
  arXiv:1012.0299. %[cond-mat.str-el].

%\cite{Eto:2011pj}
\bibitem{Eto:2011pj}
  M.~Eto, T.~Fujimori, M.~Nitta, K.~Ohashi, and N.~Sakai,
 % ``Dynamics of Non-Abelian Vortices,''
  arXiv:1105.1547.% [hep-th].



\bibitem{Sachdev:2009cp}
  S.~Sachdev, M.~A.~Metlitski, Y.~Qi, and C.~Xu,
  %``Fluctuating spin density waves in metals,''
 Phys.\ Rev.\ B {\bf  80}, 155129 (2009).
 %   [arXiv:0907.3732 [cond-mat.str-el]].


\bibitem{Hermele:2007pp}
 M.~Hermele,
  %``SU(2) gauge theory of the Hubbard model and application to the honeycomb lattice,''
  Phys.\ Rev.\ B {\bf  76}, 035125 (2007).
%  [cond-mat/0701134].

\bibitem{kr1986}
G.~Kotliar, and A.~E.~Ruckenstein,
 Phys.\ Rev.\ Lett.\  {\bf 57}, 1362 (1986);
T.~Li, P.~W\"olfle, and P.~J.~Hirschfeld,
Phys.\ Rev.\ B {\bf 40}, 6817 (1989).

\bibitem{035106}
S.~R.~Hassan, and  L.~de' ~Medici,
Phys.\ Rev.\ B {\bf 81}, 035106 (2010).




\bibitem{hep-th/0106175}
  K.~Konishi and L.~Spanu,
 % ``NonAbelian vortex and confinement,''
   Int.\ J.\ Mod.\ Phys.\ A  {\bf 18}, 249 (2003).%  [hep-th/0106175].



\bibitem{Senthil:2004cp}
T.~ Senthil, A.~Vishwanath, L.~Balents, S.~Sachdev, and M. ~P.~A. ~Fisher,
%"Deconfined" quantum critical points"
Science {\bf 303}, 1490 (2004).
% [arXiv:0311326v1 [cond-mat.str-el]].
%\cite{Sachdev:2010uz}

\bibitem{arXiv:0802.1020}
  M.~Eto, T.~Fujimori, S.~B.~Gudnason, K.~Konishi, M.~Nitta, K.~Ohashi and W.~Vinci,
  %``Constructing Non-Abelian Vortices with Arbitrary Gauge Groups,''
  Phys.\ Lett.\ B {\bf 669}, 98 (2008).% [arXiv:0802.1020 [hep-th]].

%\cite{arXiv:1007.2116}
\bibitem{arXiv:1007.2116}
  S.~B.~Gudnason, Y.~Jiang and K.~Konishi,
 % ``Non-Abelian vortex dynamics: Effective world-sheet action,''
  JHEP\ {\bf 1008}, 012 (2010).% [arXiv:1007.2116 [hep-th]].

\bibitem{Jackiw:1981ee}
  R.~Jackiw, and P.~Rossi,
  %``Zero Modes of the Vortex - Fermion System,''
  Nucl.\ Phys.\ B {\bf 190}, 681 (1981).


\bibitem{Yasui:2010yw}
  S.~Yasui, K.~Itakura and M.~Nitta,
  %``Fermion structure of non-Abelian vortices in high density QCD,''
   Phys.\ Rev.\ D {\bf 81},105003 (2010).% [arXiv:1001.3730 [math-ph]].  %%CITATION = ARXIV:1001.3730;%%
\bibitem{Roy:2010pp}
 R.~Roy,
  %``Topological Majorana and Dirac Zero Modes in Superconducting Vortex Cores,''
  Phys.\ Rev.\ Lett.  {\bf 105}, 186401(2010).
%\cite{Yasui:2010yw}
\bibitem{PhysRevB.55.1142}
  A.~Altland, and M.~R.~Zirnbauer,
    %"Nonstandard symmetry classes in mesoscopic normal-superconducting hybrid structures",
    Phys.\ Rev.\ B {\bf 55}, 1142 (1997).

\bibitem{prb195}
A.P.~Schnyder, S.~Ryu, A.~Furusaki, and A. ~W. ~W. ~Ludwig,
Phys.\ Rev.\ B  {\bf 78}, 195125 (2008).
\bibitem{cond-mat/0201547}
  E.~Babaev,
  %``Phase diagram of a planar two band superconductor: Condensation of vortices with fractional flux quantum and existence of a nonsuperconducting superfluid state in this system,''
  Nucl.\ Phys.\ B\ {\bf 686}, 397 (2004).%  [cond-mat/0201547].  %%CITATION = NUPHA,B686,397;%%

%\cite{arXiv:1010.3724}
\bibitem{arXiv:1010.3724}
  H.~Yao and D.H.~Lee,
  %``Fermionic magnons, non-Abelian spinons, and spin quantum Hall effect from an exactly solvable Kitaev Hamiltonian with SU(2) symmetry,''
  Phys.\ Rev.\ Lett.\ \ {\bf 107}, 087205 (2011).
 % [arXiv:1010.3724 [cond-mat.str-el]].
  %%CITATION = PRLTA,107,087205;%%


\bibitem{Gudnason:2009ut}
  S.~B.~Gudnason,
  %``Non-Abelian Chern-Simons vortices with generic gauge groups,''
  Nucl.\ Phys.\  B {\bf 821}, 151 (2009);
%  [arXiv:0906.0021 [hep-th]].
  %%CITATION = NUPHA,B821,151;%%
  M.~Eto and S.~B.~Gudnason,
  %``Zero-modes of Non-Abelian Solitons in Three Dimensional Gauge Theories,''
  J.\ Phys.\ AA\ {\bf 44},095401 (2011).
%  [arXiv:1009.5429 [hep-th]].
  %%CITATION = JPAGB,A44,095401;%%



\end{thebibliography}
\end{document}